\documentclass[useAMS,usenatbib]{mn2e}
\usepackage{graphics}
\usepackage{epsfig}
\usepackage{amssymb}

\title[The PPTA Sensitivity to Individual Sources of GWs]{The Sensitivity of the Parkes Pulsar Timing Array to Individual Sources of Gravitational Waves}
\author[Yardley et al.]{D. R. B. Yardley$^{1,2}$\thanks{E-mail: dyardley@physics.usyd.edu.au (DRBY)}, G. B. Hobbs$^1$, F. A. Jenet$^3$, J. P. W. Verbiest$^4$, Z. L. Wen$^5$, \newauthor R. N. Manchester$^1$, W. A. Coles$^6$, W. van Straten$^7$,
M. Bailes$^7$, N. D. R. Bhat$^7$, \newauthor S. Burke-Spolaor$^{1,7}$, D. J. Champion$^{1,8}$,  A. W. Hotan$^9$, J. M. Sarkissian$^1$ \\
$^{1}$Australia Telescope National Facility, CSIRO, P.O. Box 76, Epping, NSW 1710, Australia.\\
$^2$Sydney Institute for Astronomy, School of Physics A29, The University of Sydney, NSW 2006, Australia.\\
$^3$Center for Gravitational Wave Astronomy, University of Texas at Brownsville, 80 Fort Brown, Brownsville, TX 78520, USA. \\
$^4$Department of Physics, West Virginia University, Morgantown, WV 26506, USA. \\
$^5$National Astronomical Observatories, Chinese Academy of Sciences, Jia-20 DaTun Road, ChaoYang District, Beijing 100012, China. \\
$^6$Electrical and Computer Engineering, University of California at San Diego, La Jolla, California, USA. \\
$^7$Centre for Astrophysics and Supercomputing, Swinburne University of Technology, P.O. Box 218, Hawthorn VIC 3122, Australia. \\
$^8$ Max-Planck-Institut f\"ur Radioastronomie, Auf Dem H\"ugel 69, 53121, Bonn, Germany.\\
$^9$Department of Imaging and Applied Physics, Curtin University, Bentley, Western Australia, Australia. \\
}
\begin{document}
\date{Accepted 2010 May 2. Received 2010 April 20; in original form 2009 November 23}

\pagerange{\pageref{firstpage}--\pageref{lastpage}} \pubyear{2009}

\maketitle

\label{firstpage}

\begin{abstract}
We present the sensitivity of the Parkes Pulsar Timing Array to gravitational waves emitted by individual super-massive black-hole binary systems in the early phases of coalescing at the cores of merged galaxies. Our analysis includes a detailed study of the effects of fitting a pulsar timing model to non-white timing residuals. Pulsar timing is sensitive at nanoHertz frequencies and hence complementary to LIGO and LISA. We place a sky-averaged constraint on the merger rate of nearby ($z < 0.6$) black-hole binaries in the early phases of coalescence with a chirp mass of $10^{10}\,\rmn{M}_\odot$ of less than one merger every seven years. The prospects for future gravitational-wave astronomy of this type with the proposed Square Kilometre Array telescope are discussed.
\end{abstract}


\section{Introduction} \label{sec:intro}

In the era of ground- and space-based gravitational-wave (GW) detectors, GW astronomy is becoming increasingly important for the wider astronomy and physics communities. The ability of the current GW community to provide either limits on, or detections of, GW emission is of enormous importance in characterising astrophysical sources of interest for further investigation. It is possible that GW detection will provide the only means to probe some of these sources. The sensitivity of existing and future observatories to individual GW sources, such as neutron-star binary systems and coalescing black-hole binary systems, has been calculated in the $\sim$\,kHz and $\sim$\,mHz frequency ranges. The sensitivity curves of the Laser Interferometer Gravitational-Wave Observatory \citep{2009PhRvD..80j2001A}\footnote{See http://www.ligo.caltech.edu/advLIGO/}, Virgo \citep{2006CQGra..23S..63A}\footnote{See http://www.virgo.infn.it/} and the Laser Interferometer Space Antenna \citep[LISA: ][]{2000PhRvD..62f2001L}\footnote{See http://lisa.nasa.gov/} cover these frequency ranges. The sensitivity of GW detectors to individual sources of GWs at lower frequencies has not yet been presented in detail.

Radio observations of pulsars have long been proposed as a means of detecting low-frequency GWs \citep{1978SvA....22...36S, det79,1983ApJ...265L..39H, 2003ApJ...590..691W, 2005ApJ...625L.123J}. Pulsars are used as a GW detector via comparison between a model for their pulse arrival times and high precision measurements of these ``times-of-arrival'' (TOAs) at a radio telescope over a period of years \citep[see, e.g.,][]{2004hpa..book.....L, 2006MNRAS.372.1549E}. Pulsar timing is most sensitive to GWs in the $\sim$\,nHz frequency range \citep[see, e.g.,][]{2005ApJ...625L.123J, hjl+09}. The sources most likely to produce a detectable GW signal in this frequency range are super-massive black-hole binary systems (SMBHBs) in the early phases of coalescence at the cores of merged galaxies \citep[see, e.g.,][]{2009MNRAS.394.2255S}.

Earlier work \citep{1983ApJ...265L..35R, 1994ApJ...428..713K, lom02, 2006ApJ...653.1571J} aimed to limit the amplitude of the stochastic background of GWs, either using observations of an individual pulsar, or using precise and contemporaneous timing of several pulsars (a ``pulsar timing array'' [PTA] ). Here we describe methods to limit or detect individual GW sources. In \citet{Wen09}, the non-detection of GWs from a single SMBHB in the pulsar timing observations presented in \citet{2006ApJ...653.1571J} was used to place limits on the coalescence rate of SMBHBs with a range of redshifts and chirp masses. The timing residuals reported in \citet{2006ApJ...653.1571J} were carefully selected because the power spectrum of each pulsar was consistent with ``white noise'' -- that is, the spectral power is statistically constant across all frequencies. However, the timing residuals of most pulsars (including some in our sample) are not consistent with white noise; rather they are affected by a variety of phenomena including changes in the interstellar medium \citep{yhc+07}, calibration errors \citep{van06} and irregular spindown behaviour known as ``timing noise'' \citep{2010MNRAS.402.1027H}. 

Previous authors \citep{2001ApJ...562..297L, 2004ApJ...606..799J} have addressed the issue of characterising the GW signals in pulsar timing residuals expected from SMBHBs in the early phase of coalescing. However, both works considered very specific applications to known astrophysical systems, namely the GWs being emitted from the radio galaxy 3C66B or the Galactic Centre (Sagittarius A*) and nearby massive dark objects. \citet{2001ApJ...562..297L} showed that the maximum possible induced timing residual caused by a binary black hole in Sagittarius A* is around 14\,ns, which is below current limits. \citet{2004ApJ...606..799J} showed that they could rule out a proposed SMBHB system at the core of 3C66B \citep{2003Sci...300.1263S} with 95\% confidence using a publicly available pulsar dataset \citep{1994ApJ...428..713K}.

The main aim of this paper is to calculate the sensitivity of pulsar timing data to individual sources of sinusoidal GWs. The analysis takes account of all the issues affecting these data, including fitting of pulsar parameters, small amounts of non-white noise and sampling effects. This paper is organised as follows: \S~\ref{sec:obsns} describes the observations used to produce the sensitivity curves, \S~\ref{sec:method} describes our method for detecting significant sinusoids in pulsar timing residuals, \S~\ref{sec:disc} gives our results and describes some implications and \S~\ref{sec:conc} concludes the paper. The Appendix contains extra details of the detection technique we have developed. This includes a more thorough description of the issues encountered, in particular those caused by fitting for pulsar parameters and the irregular sampling of the timing residuals.

\section{Observations} \label{sec:obsns}

\begin{table*}
 \centering
 \begin{minipage}{180mm}
  \caption{Eighteen of the Parkes Pulsar Timing Array pulsars and their rms timing residuals from the dataset presented in \citet{vbc+09}.} \label{tbl:jorisData}
  \begin{tabular}{@{}ccccccc@{}}
  \hline
    PSRJ & Period & DM & $P_b$ & Weighted RMS & Span & No. of \\
    & (ms) & (cm$^{-3}$pc) & (d) &  Residual ($\mu$s) & (years) & Observations\\
 \hline
J0437$-$4715 & 5.757 & 2.65 & 5.74 & 0.20 & 9.9 & 2847\\
J0613$-$0200 & 3.062 & 38.78 & 1.20 & 1.56 & 8.2 & 190\\
J0711$-$6830 & 5.491 & 18.41 & -- & 3.23 & 14.2 & 227\\
J1022+1001 & 16.453 & 10.25 & 7.81 & 1.62 & 5.1 & 260\\
J1024$-$0719 & 5.162 & 6.49 & -- & 4.22 & 12.1 & 269\\
J1045$-$4509 & 7.474 & 58.15 & 4.08 & 6.64 & 14.1 & 401\\
J1600$-$3053 & 3.598 & 52.19 & 14.34 & 1.14 & 6.8 & 477\\
J1603$-$7202 & 14.842 & 38.05 & 6.31 & 1.92 & 12.4 & 212\\
J1643$-$1224 & 4.622 & 62.41 & 147.02 & 2.50 & 14.0 & 241\\
J1713+0747 & 4.570 & 15.99 & 67.83 & 0.20 & 14.0 & 392\\
J1730$-$2304 & 8.123 & 9.61 & -- & 2.51 & 14.0 & 180\\
J1732$-$5049 & 5.313 & 56.84 & 5.26 & 3.24 & 6.8 & 129\\
J1744$-$1134 & 4.075 & 3.14 & -- & 0.62 & 13.2 & 342\\
J1857+0943 & 5.362 & 13.31 & 12.33 & 1.21 & 22.1\footnote{There is a gap of $\sim$11\,years between the end of the data presented in \citet{1994ApJ...428..713K} and the beginning of data collected with the Parkes telescope.} & 376\\
J1909$-$3744 & 2.947 & 10.39 & 1.53 & 0.17 & 5.2 & 893\\
J2124$-$3358 & 4.931 & 4.62 & -- & 4.03 & 13.8 & 416\\
J2129$-$5721 & 3.726 & 31.85 & 6.63 & 2.19 & 12.5 & 179\\
J2145$-$0750 & 16.052 & 9.00 & 6.84 & 1.82  & 13.8 & 377\\
\hline
\end{tabular}
\end{minipage}
\end{table*}

The millisecond pulsar timing process usually consists of using a large-aperture telescope to observe a particular pulsar and forming a mean pulse profile using an ephemeris to fold the incoming data at the correct apparent period. Since, on average, millisecond pulsar profiles are largely invariant, a shift between the standard template pulse profile and the observed profile can be established, leading to a site arrival time. After transforming to the arrival time at the Solar System barycentre we obtain a barycentric TOA. After clock corrections are applied, and a pulsar model is fitted we obtain a timing residual. Timing residuals are generally dominated by noise, but may contain systematic errors induced by our instrumentation and more subtle effects such as those induced by GWs.

The observations used in this analysis were published by \cite{vbv+08} and \cite{vbc+09}, who presented results from observations of 20 pulsars using the Parkes radio telescope\footnote{The first eight years of TOAs for PSR J1857+0943 are obtained from publicly available data collected using the Arecibo radio telescope and presented in \citet{1994ApJ...428..713K}.}. Many of these pulsars exhibit some low-frequency noise in their timing residuals, which must be accounted for in our analysis. For two of the pulsars, PSRs J1824$-$2452 and J1939+2134, the residuals are dominated by low-frequency noise that complicates the spectral analysis procedures for little gain in sensitivity. Hence we remove them from our sample.

The pulsars have been timed with a weighted rms residual of $\sim 0.2 -  7\,\mu$s for a period of $\sim 10$ years. The specifications of each set of timing residuals are given in Table \ref{tbl:jorisData}, where, in column order, we present the pulsar name in the J2000 coordinate system, pulse period, dispersion measure, orbital period, weighted rms residual, data-span and number of recorded TOAs. For full details of TOA estimation and data processing, see \citet{vbc+09}. The timing residuals for the full set of 20 pulsars are shown in Figure 1 of \citet{vbc+09}.

All the observations were made in the 20\,cm (1.4\,GHz) band, except for PSR J0613$-$0200 for which a timing solution was obtained in the 50\,cm (685\,MHz) band. Observations between 1994 and November 2002 were made with either one or two 128\,MHz-wide bands, but these data varied greatly in quality. Observations after November 2002 were taken with a phase-coherent dedispersion system, the Caltech-Parkes-Swinburne-Recorder-2 \citep[CPSR2; see][]{2003ASPC..302...57B}, over two 64\,MHz-wide observing bands centred at 1341\,MHz and 1405\,MHz. The typical observation length was 1 hour.

\section{Method} \label{sec:method}

\subsection{GW-Induced Timing Residuals}

\begin{figure*}
\centering
\epsfig{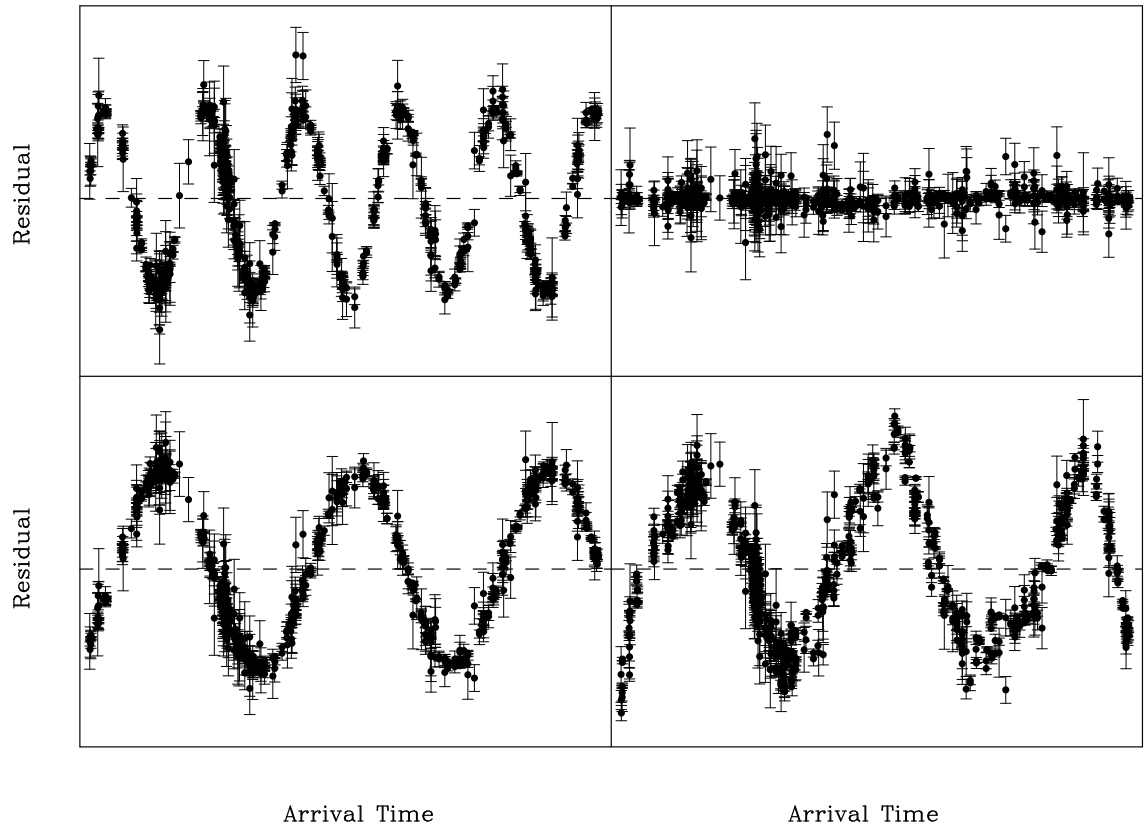}
\vspace{0.4cm}
\caption{
Attenuation of timing signals caused by pulsar parameter fitting. In each plot the abscissa is the centred MJD and the ordinate gives the magnitude of each timing residual; all plots have the same scale. The dotted lines indicate zero residual. The data plotted are formed by adding a simulated GW signal to the timing residuals for PSR J1909$-$3744 which are described in Section \ref{sec:obsns} and in \citet{vbc+09}. The top row shows a GW signal with a period of one year (top left) being completely removed after fitting (top right). The bottom row shows a GW signal with a period of two years (bottom left) being largely unaffected by the fitting procedure (bottom right). These figures were produced using simulated data from the pulsar timing package \textsc{Tempo2} \citep{hem06, 2006MNRAS.372.1549E, hjl+09}.
}
\label{fig:signals}
\end{figure*}

The timing residual induced by a GW is the integral of the interaction between the GW and the electromagnetic pulses emitted by the pulsar over the path of the pulses. For our analysis, we assume that the GW is emitted by a binary system in a circular orbit. For an equal-mass binary, the lifetime of an SMBHB scales as \citep[adapted from][]{2001ApJ...562..297L}
\begin{equation}
\label{eq:lifetime}
\tau = 2.2 \times 10^4\,\rmn{yr} \left(\frac{M}{10^9\,M_\odot}\right)^{-5/3}\left(\frac{P_\rmn{orb}}{730\,\rmn{days}}\right)^{8/3}
\end{equation}
where $M$ is the total mass of the system and $P_\rmn{orb}$ is the orbital period\footnote{Note that $2P_\rmn{GW} = P_\rmn{orb}$, where $P_\rmn{GW}$ is the period of the emitted GWs.}. For a SMBHB with $M=10^9\,\rmn{M}_\odot$ and $P_\rmn{orb} = 730\,\rmn{d}$ (which would emit GWs with a one year period), the lifetime is three orders of magnitude larger than the typical data span of pulsar timing observations. This means no significant chirping of the GW signal will occur over the duration of the observations. Therefore we can calculate analytically, using Equations (1) $-$ (7) of Jenet et al. (2004), the expected GW signal in the timing residuals. The result is that the induced timing residuals will contain two sinusoidal signals which can be called the ``Earth term'' and the ``pulsar term''.

However, evolution of the SMBHB is sometimes a measurable effect over the timescale of the light travel time from the pulsar to Earth as, for example, in the evolution of the proposed SMBHB in 3C66B \citep{2004ApJ...606..799J} which results in two distinct periodicities in the timing residuals. In this work we ignore this longer timescale evolution, so the GW-induced quadrupolar space-time distortions at the Earth and the pulsar will always have the same frequency. However we have allowed the two periodicities to be offset in phase which can alter the amplitude of the signal in the timing residuals. We hence reduce the problem of detecting GW emission from a non-evolving circular binary system to identifying the presence of a significant sinusoid in the timing residuals. To confirm that a significant sinusoid in the timing residuals of a given pulsar is caused by GWs, one would need to ensure that the expected signature of the GW \citep[see, e.g.,][]{det79} is present in the timing residuals of other pulsars.

To determine the residuals a particular SMBHB will induce in our data, we begin with the expected GW strain\footnote{By convention $h_c$ is the strain from a GW background while $h_s$ gives the GW strain from a single source.} emitted by a single SMBHB \citep{tho87}:
\begin{equation}
\label{eq:SMBHBh_c}
h_s = 4\sqrt{\frac{2}{5}}\frac{\left(GM_c\right)^{5/3}}{c^4D(z)}    \left[\pi f\left(1 + z\right)\right]^{2/3}
\end{equation}
where $M_c = \left(M_1M_2\right)^{3/5}\left(M_1+M_2\right)^{-1/5}$ is the chirp mass of the SMBHB with member masses $M_1$ and $M_2$, $G$ the gravitational constant, $c$ the vacuum speed of light, $f$ the observed GW frequency (which is in general different to the emitted frequency), $z$ the redshift of the SMBHB and $D(z)$ is the comoving distance to the SMBHB, given by
\begin{equation}
\label{ }
D(z) = \frac{c}{H_0}\int_0^z\frac{dz'}{E(z')}
\end{equation}
where $H_0$ is Hubble's  constant, taken to be $72$\,km\,s$^{-1}$\,Mpc$^{-1}$ and $E(z) = H(z) / H_0 = \sqrt{\Omega_\Lambda + \Omega_m(1+z)^3}$ under a $\Lambda$CDM cosmological model \citep[see, e.g.,][]{2009PhRvD..80f7501S}. For this work we assume  $\Omega_\Lambda = 0.7$ \citep[see, e.g.,][]{2009ApJS..180..330K} giving $\Omega_m = 0.3$.

A non-evolving GW source will induce a sinusoidal variation in the pulsar timing residuals, with amplitude \citep{Wen09}
\begin{equation}
\label{eq:ampres}
A_{\rm res} = \frac{h_s}{\omega}(1+\cos\theta)\sin(2\phi)    \sin\left[\frac{\omega D_p(1-\cos\theta)}{2c}\right] 
\end{equation}
where $\omega = 2\pi / P_\rmn{GW}$ is the GW frequency in rad $\mathrm{s}^{-1}$, $\theta$ is the angle between the direction from which the GWs emanate and a vector from the Earth to the pulsar, $\phi$ is the GW polarisation angle and $D_p$ is the distance to the pulsar.

An important feature of any pulsar timing analysis is the process of parameter determination for the model of the pulsar. This process is equivalent to fitting out a range of signals from the time series of residuals. Figure \ref{fig:signals} shows the effect this can have on GW detection -- a GW signal with a period of one year (top left panel) will be almost completely removed after fitting (top right panel) because this signal mimics an error in the pulsar position. However, a GW signal with a period of two years (bottom left) is only slightly attenuated by fitting (bottom right). To determine the post-fit timing residuals, we add the effect of a sinusoidal GW point source directly to the TOAs using \textsc{Tempo2} \citep{hem06,hjl+09} and then perform the standard pulsar timing fitting procedure on these modified TOAs.

\subsection{Producing the Sensitivity Curve}\label{sec:tech}

The detection of a sine wave in the presence of noise with known statistics is a well-studied problem with a simple optimal solution, the maximum likelihood estimator. A number of algorithms can be used, depending on the characteristics of the data. In this case the problem is complicated by the fact that the data are irregularly sampled and the noise consists of at least two components. The noise has a white component which varies from sample to sample; this component is well understood and we have a variance estimate for the white noise at each sample point. The noise also has a non-white component for which the source is unknown. We assume that it has a smoothly varying power spectrum and attempt to estimate this from the data.

We use one of the most common spectral estimation tools: an unweighted Lomb-Scargle periodogram \citep{1992nrca.book.....P}. By ``unweighted'', we mean that the individual TOA errors are not taken into account when calculating the power spectrum. The Lomb-Scargle periodogram technique is not valid for datasets which exhibit a steeply sloping spectrum (the timing residuals of some young pulsars exhibit very high power levels only at low frequencies). While many of our timing residuals do not exhibit a flat Lomb-Scargle periodogram, none of our 18 datasets has a sufficiently steeply sloping spectrum to invalidate this approach. It has been argued \citep[see][]{1999ApJ...526..890C} that a `floating mean' periodogram should be used to obtain the correct spectral estimates when detecting long period signals in sparsely and unevenly sampled data; this potential improvement will be addressed in a future paper. We briefly describe our approach for producing a sensitivity curve here; full details are provided in the Appendix. 

To make a detection of a significant sine wave in our timing residuals, we make a simple model of the noise in the Lomb-Scargle periodogram of the timing residuals and use this model to define a set of detection thresholds. These thresholds are set high enough that the probability of recording a detection at any frequency across the entire observed power spectrum when no signal is present is 1\% (the ``false alarm probability''). 

We then inject simulated GW signals of known frequency with random polarisations, random sky locations and at a range of strain amplitudes to determine the strain that gives a detection in the corresponding frequency channel in 95\% of the simulations. The GW-induced quadrupolar space-time distortions at the Earth and the pulsar can interfere constructively or destructively. This process gives the sky- and polarisation-averaged sensitivity as a function of GW frequency over the range $f \sim \left(10\,\rmn{years}\right)^{-1}$ to $f \sim \left(1\,\rmn{month}\right)^{-1}$.

There are two aspects to this detection strategy, namely the false alarm probability (1\%) and the probability of making a detection (95\%). Using a false alarm probability of 1\% means that in a particular simulated dataset, any detection made will be a 3-$\sigma$ detection. However, the sensitivity of a pulsar GW detector is not the same for every possible GW source; for instance, a single pulsar cannot be used to detect GWs propagating along the line of sight from the Earth to the pulsar. Our sensitivity curves give the GW amplitude at which the probability of making a 3-$\sigma$ detection at a random position on the sky and with a random polarisation is 95\%. If the GW polarisation and the pulsar-Earth-source angle were favourable (e.g. $\theta = \pi/2$ and $\phi = \pi/4$ in equation \ref{eq:ampres}), then for a single pulsar an improvement by a factor of $\sim 10-15$ in sensitivity could be achieved compared with the sky-averaged sensitivity. One of the advantages of timing an array of pulsars to detect sinusoidal GWs is that there is a significantly smaller region of sky in which a full array has low sensitivity.

We are interested in answering two questions. The first is ``What is the largest GW source at a particular frequency that could be present in the measured timing residuals?'' This will give an upper bound on the amplitude of individual GW sources in our data at that frequency. This question is answered by comparing simulated GW sources to our observed timing residuals. We simulate GW sources at a given frequency at random sky locations and adjust the amplitude of these sources until the power of the GW sinusoid exceeds the power in the observed timing residuals at that frequency in 95\% of simulations. This approach gives the most conservative upper limit, since we are allowing for the possibility that {\it all} the power we observe at this frequency results from one sinusoidal GW point source. We will determine this upper limit for our datasets in \S~4.

The second question is ``If there were a GW source with a particular frequency somewhere on the sky, what is the minimum strain amplitude that would produce a detectable signal at that frequency in our dataset?" In order to answer this we add simulated sinusoidal GW signals to our TOAs, perform the standard pulsar timing analysis and calculate the minimum amplitude at which we would detect a significant sinusoid at the input GW frequency in our data {\it if we had collected that dataset at a telescope}. This means we must account for all the sources of noise in our pulsar detector. The threshold for detection at any frequency across the observed power spectrum will often be $\sim$3 times greater than the locally-averaged power level. This gives our sensitivity to detecting these sinusoids, rather than just limiting their amplitude. For very large amplitude sine waves with period $\gtrsim T_\rmn{obs}$, a signal will often be detectable at a slightly higher frequency than the input frequency because we can detect the side lobes of the large input signal. We have not allowed detections at different frequencies to the input GW frequency in our implementation. The sensitivity curve corresponding to our datasets is derived in \S~4.

The periodogram frequency range is from $\frac{1}{T_\rmn{obs} }$ to $\frac{N_\rmn{pts}}{2T_\rmn{obs}}$ for a single pulsar (where $T_\rmn{obs}$ is the time-span of the observations and $N_\rmn{pts}$ is the number of timing residuals for that pulsar). Note that $\frac{N_\rmn{pts}}{2T_\rmn{obs}}$ would be the Nyquist frequency for that pulsar if its timing residuals were regularly sampled. If we have multiple pulsars then we can perform a weighted sum of their power spectra to increase our sensitivity. For this work we first make a simple frequency-dependent model of the noise in the pulsar power spectrum and then weight each pulsar by the inverse of the noise model for that pulsar. For the case of spectrally white timing residuals, the noise model will be independent of frequency, so this is equivalent to weighting by the inverse variance of the timing residuals. To perform the sum requires the use of a common frequency gridding, so when analysing multiple pulsars the periodogram ranges from $(30\,\rmn{years})^{-1}$ to $(4\,\rmn{weeks})^{-1}$ for all pulsars.

Our detection technique is straightforward to implement, but for many pulsars with differing data-spans and noise properties etc., it will not be optimal. After adding a simulated GW signal to each dataset, some of the power at the frequency of the GW signal will be leaked into adjacent channels, meaning that the noise model will be higher near the GW frequency, leading to fewer detections. We also use a simple weighting scheme to combine multiple pulsars which gives a factor of $\sim 5$ improvement over a simple, non-weighted addition of the power spectra for the different pulsars. However, the exact weighting used in an incoherent detection scheme such as ours does not significantly change the overall sensitivity of the array. A small improvement in sensitivity may also be gained by allowing for the evolution of the GW source over the light-travel-time from the pulsar to the Earth and then searching for a two-frequency response in each pulsar's power spectrum. The 18-pulsar array sensitivity could also be improved by `phasing up' the timing array to enable a coherent sum of the GW signal in each dataset. However, such a detection scheme is considerably more complex and will be addressed in a future paper.

\section{Results and Discussion} \label{sec:disc}

In this section we present the sensitivity of the Parkes Pulsar Timing Array (PPTA) to sinusoidal GW sources using the dataset described in Table \ref{tbl:jorisData} and account for all the observed features in the sensitivity curves. We also describe some of the implications of the non-detection of sinusoidal GWs in our dataset and give predictions for a future timing array project using the SKA.

\subsection{The sensitivity using some individual pulsars}

\begin{figure*}
\centering
\epsfig{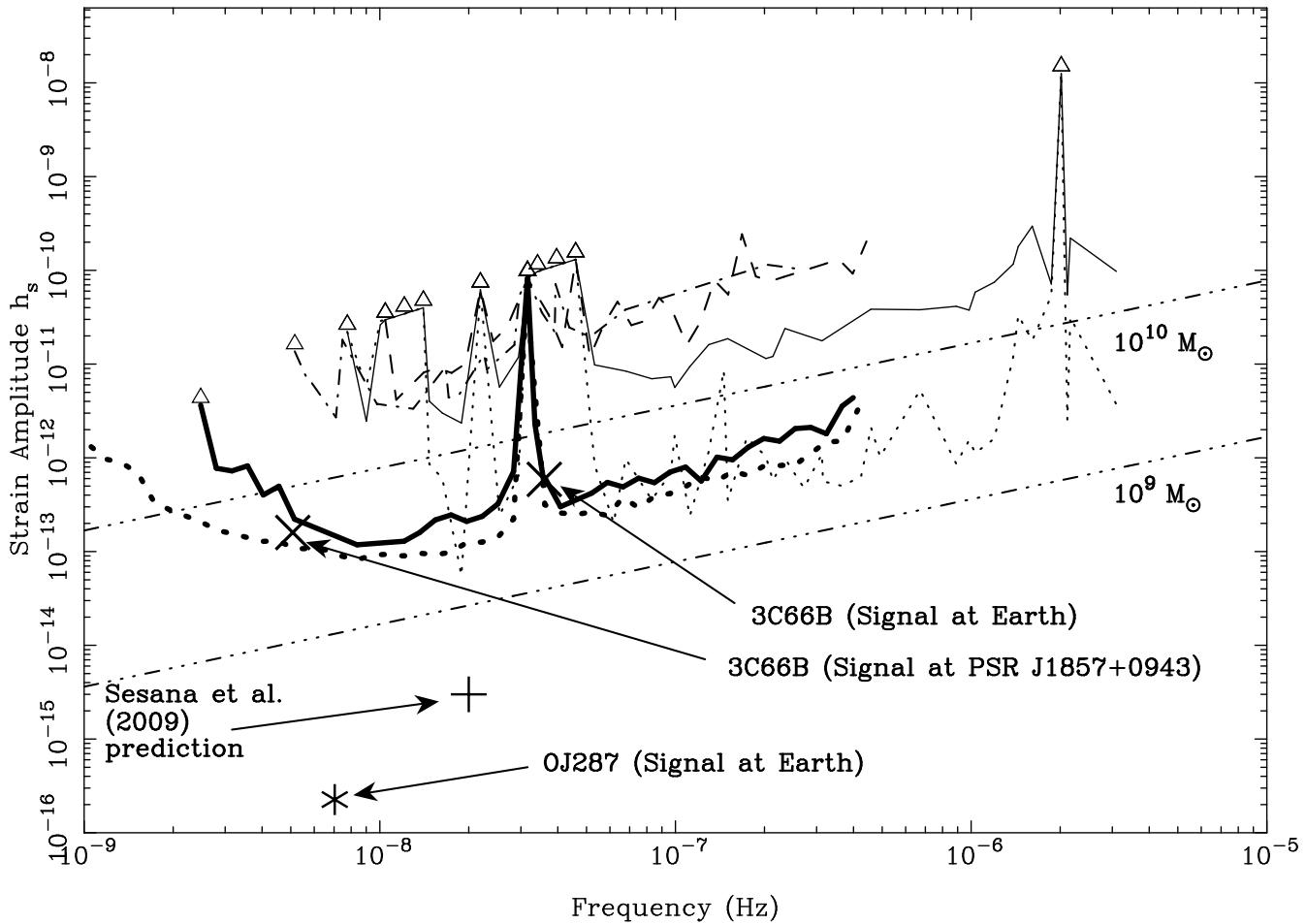}
\caption{
Sensitivity curves for PSRs J0437$-$4715 (thin solid line), J1713+0747 (dashed), J1857+0943 (dot-dashed) and the 18-pulsar timing array assuming an incoherent detection scheme is used (thick solid line). The abscissa gives the GW frequency, the ordinate gives the minimum detectable strain amplitude of a sinusoidal GW point source with a random polarisation, phase and sky-position. The thin dotted line is the maximum attainable sensitivity using PSR J0437$-$4715 assuming optimal sky-location and polarisation of the GW source. An open triangle indicates that the plotted value is in fact a lower bound on the detectable amplitude at that frequency. The straight triple-dot-dashed lines indicate the expected signal from an individual SMBHB with equal member masses of $10^{9}$M$_\odot$ or $10^{10}$M$_\odot$ if it were located at the mean distance of the Virgo cluster. The `$\times$' symbols are the expected signals at the Earth now and at PSR J1857+0943 $\sim$2700 years ago caused by the proposed SMBHB at the core of the radio galaxy 3C66B. The `$*$' symbol is the expected signal caused by the proposed SMBHB at the core of OJ287. The `+' symbol is the GW strain and frequency emitted by a typical resolvable SMBHB as plotted in Figure 2 of \citet{2009MNRAS.394.2255S}. Also shown on the plot is the upper limit on the amplitude of sinusoidal GW point sources as a function of frequency using the 18-pulsar timing array (thick dotted line); in this case the ordinate gives the maximum amplitude GW source which could be present in our data.
}
\label{fig:SensCurve}
\end{figure*}

In Figure \ref{fig:SensCurve} we plot the sky- and polarisation-averaged sensitivity curves for PSRs J0437$-$4715 (thin solid line), J1713+0747 (dashed line) and J1857+0943 (dot-dashed line) where each pulsar has been analysed individually. The open triangles on the plot indicate that the plotted ``detectable'' amplitude at that frequency value is a lower bound. The thin dotted line indicates the sensitivity of PSR J0437$-$4715 to a hypothetical SMBHB located at a right ascension of 4$^\rmn{h}$37$^\rmn{m}$ and a declination of +42$^\circ$45$^\rmn{m}$ and emitting purely `plus' polarised GWs. This line indicates the maximum sensitivity obtainable with this dataset if the GW source position and polarisation were favourable in every simulation. The ratio of this thin dotted line to the thin solid line gives the factor of $\sim 10 - 15$ improvement in sensitivity for optimal sky-location and polarisation discussed in \S~\ref{sec:tech}. Also shown are the expected signals at a range of frequencies from two hypothetical SMBHB systems at the mean distance of the Virgo cluster \citep[taken to be 16.5\,Mpc, from][]{2007ApJ...655..144M}, with equal member masses of $10^{9}$M$_\odot$ or $10^{10}$M$_\odot$.

The reduction in sensitivity caused by fitting for the pulsar's position will be at the same frequency of $(1\,\rmn{yr})^{-1}$ for all pulsars. However, fits for orbital parameters will also reduce sensitivity to GWs, but at different frequencies for each pulsar.  All pulsars exhibit a reduction in sensitivity at low frequencies; this is caused by the fit of a quadratic polynomial to the TOAs required to model the pulsar spin-down, as well as the fitting of ``jumps'' to many of the datasets to connect the timing residuals obtained with different backend systems (see below). 

As the GW frequency increases, the strength of the signal in our residuals becomes weaker for a given strain, as described by equation (\ref{eq:ampres}). At the highest frequencies, our sensitivity is limited by the sampling of the timing residuals. This is particularly evident in the sensitivity curve for the 18-pulsar timing array where there is a turn-up in the sensitivity curve at the last few frequency values corresponding to a decrease in sensitivity there.

The sensitivity of our detection technique to low-frequency sinusoidal GWs in irregularly-sampled data (where the GW period is similar to the data-span) is reduced compared to treating regularly-sampled data since there is no clear way to distinguish between the excess low-frequency noise seen in many millisecond pulsars and spectral leakage from the low-frequency GWs. Some pulsars in our sample do not exhibit excess low-frequency noise (e.g., PSR J1857+0943), meaning that the power spectrum with no GWs added may be modelled with a constant. However, as soon as a low-frequency sinusoidal GW source is added to these residuals, leakage from the low-frequency signal is difficult to distinguish in our technique from standard pulsar timing noise and interstellar medium variations, so our model of the power spectrum must account for this confusion. In a regularly-sampled time series with weak red noise, spectral leakage is less severe and thus there is no such confusion.

In the sensitivity curve for PSR J0437$-$4715 there is a loss of sensitivity at a frequency of $(540\,\rmn{days})^{-1}$, or $\sim$21\,nHz. This is caused by the fitting of several constant time offsets between the data collected using different observing backend systems; such offset fits absorb GW power at low frequencies. If overlapping data exist between the different observing backends, these offsets can be precisely determined and held fixed in subsequent processing. Even if no overlapping data exist, it is sometimes possible to eliminate these arbitrary offsets without losing phase connection in the timing solution. Our analysis takes into account all of the offsets fitted by \citet{vbc+09}. There is also a loss in sensitivity just above the $(1 \rmn{yr})^{-1}$ frequency for this pulsar. This is caused by the sampling of the dataset.

\subsection{The sensitivity of the PPTA and some likely single sources}\label{sec:PPTAsens}

Figure \ref{fig:SensCurve} contains the sky-averaged sensitivity attainable for the 18 pulsars in our dataset assuming an incoherent detection scheme is used and the GW source position and polarisation are unknown. The plotted frequency range $(30\,\rmn{years})^{-1}$ -- $(4\,\rmn{weeks})^{-1}$ is chosen to demonstrate the high- and low-frequency sensitivity limits for our pulsar timing datasets. At the lowest frequencies, our sensitivity is limited by the fact that our longest dataset is much shorter than 30\,years and by the necessary period derivative and jump fits. At the highest frequencies we are limited by the sampling of our timing residuals; that is, $(4\,\rmn{weeks})^{-1}$, the nominal Nyquist frequency for the PPTA.

Figure \ref{fig:SensCurve} also shows the upper limit attainable using the 18 pulsars from the \cite{vbc+09} dataset (the thick dotted line at the bottom). This limit curve was obtained with 95\% confidence as described in Section \ref{sec:tech} and in the Appendix. For some pulsars a different-order polynomial model to the detection case was chosen in order to accurately model the power spectrum with no GWs added. \citet{2001ApJ...562..297L} placed a 99\% confidence limit showing that they could rule out signal amplitudes as small as 150\,ns in their residuals at a period of 53\,days, corresponding to SMBHB orbital periods of 106\,days. Using our longer datasets and the same 99\% confidence level, we can place a better limit of around 120\,ns at this frequency. At signal periods of 1000\,days where some of our datasets exhibit excess low-frequency noise, we obtain a 99\% confidence limit of 190\,ns which is worse than the \citet{2001ApJ...562..297L} limit of 170\,ns. However, there is no evidence that their analysis takes into account the effects of red noise present in their residuals.

The two `$\times$' symbols in Figure \ref{fig:SensCurve} indicate the expected strain amplitude and frequency of the proposed SMBHB at the core of the radio galaxy 3C66B \citep{2003Sci...300.1263S}. In order to determine the expected strain amplitude, we use equation (\ref{eq:SMBHBh_c}) with the redshift and masses given in the original paper ($m_1 = 4.91\times 10^{10}\,\rmn{M}_\odot, m_2 = 4.91\times 10^9\,\rmn{M}_\odot, z=0.0215$) and a distance to the source of $90$\,Mpc, implied by the low-redshift distance approximation $D = cz / H_0$. The frequencies of the signal at the Earth and at PSR J1857+0943 ($f_\rmn{Earth} = 1 / 0.88\rmn{yr}, f_\rmn{J1857+0943} = 1 / 6.24\rmn{yr}$) were obtained from \cite{2004ApJ...606..799J}. This system was ruled out with 95\% confidence by \citet{2004ApJ...606..799J}. Our results show that even with a blind search of the Verbiest et al. data, where we know neither the sky position nor the frequency of the GWs, we would detect the GW-induced oscillations at the Earth caused by this source. The expected signal is well below the plotted sensitivity curve for PSR J1857+0943 even though \citet{2004ApJ...606..799J} only used the publicly available timing residuals for PSR J1857+0943. However, their technique is analogous to our limit technique, whereas the sensitivity curve plotted for PSR J1857+0943 in Figure \ref{fig:SensCurve} assumes we are aiming to {\it detect} such sources of GWs. Furthermore, our sensitivity curve is sky-averaged whereas they used the known position and frequency of the proposed GW source in their analysis (by chance it had a very favourable sky-location with an angle of 81.5$^\circ$ between the Earth-pulsar vector and the Earth-3C66B vector).\footnote{\citet{2004ApJ...606..799J} also underestimated the distance to the proposed GW source in 3C66B by around 10\%.}

The `$*$' symbol in Figure \ref{fig:SensCurve} indicates the expected GW strain and frequency for the candidate SMBHB in the blazar OJ287. A $\sim$12\,yr-periodic signal has been identified in its optical outbursts \citep{1996A&A...305L..17S}, but other parameters of the system are not well-constrained. We parametrise the SMBHB as follows: member masses $1.3\times 10^8\,\rmn{M}_\odot$ and $1.8\times 10^{10}\,\rmn{M}_\odot$, orbital period 9 years (observed GW period $4.5$ years), eccentricity 0\footnote{In \citet{2009ApJ...698..781V} the eccentricity is estimated to be 0.7, but we do not consider eccentric SMBHBs in this paper.}, redshift 0.306, distance $1.3$\,Gpc. The distance was again obtained using $D = cz / H_0$, which is an acceptable approximation given the imprecision in the other parameter measurements and the fairly low redshift of this system \citep[see footnote 1 in][]{2004PASA...21...97D}. The GW signals emitted by this system induce timing residuals of around 5\,ns which are well below current limits.

In \citet{2008MNRAS.390..192S} a study was presented of the generation of the stochastic gravitational-wave background from the cosmic population of SMBHBs. This work showed that the stochastic background of GWs is likely to be detected using a pulsar timing array in the near future. In \citet{2009MNRAS.394.2255S} the individual resolvable SMBHBs were considered. They predicted that at least one SMBHB will induce timing residuals around 5 $-$ 50\,ns, which is below our current sensitivity. We choose (from the upper left panel of their Figure 2) a representative resolvable single source from their simulations, with an emitted GW frequency of $2\times 10^{-8}$\,Hz and a characteristic induced timing residual of 25\,ns. The signal from this source is indicated by the `+' symbol in Figure \ref{fig:SensCurve}. This is a typical resolvable SMBHB, thus it is likely that several sources will emit GWs with a larger amplitude than this. We emphasise that we do not yet have long data-spans with sufficiently low rms residual to detect such sources. A stochastic background of GWs may be detected using a pulsar timing array within the next few years.

\begin{figure}
\centering
\epsfig{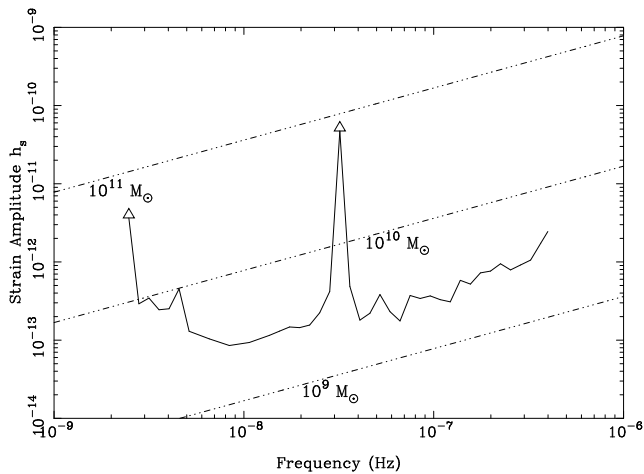}
\caption{
Sensitivity of the PPTA using the 18-pulsar Verbiest et al. dataset for detecting signals from SMBHBs located at the sky-position and mean distance of the Virgo cluster. The abscissa gives the GW frequency, the ordinate gives the minimum detectable strain amplitude of a sinusoidal GW point source emanating from the direction of the mean sky-position of the Virgo cluster with a random polarisation and phase. The open triangles indicate that the plotted value is a lower bound on the detectable amplitude at those frequencies. The dot-dashed lines indicate the expected signals from three different types of SMBHB if they were located in the Virgo cluster, with equal member masses $10^{9}$M$_\odot$, $10^{10}$M$_\odot$ and $10^{11}$M$_\odot$ as labelled. 
}
\label{fig:VIRGO}
\end{figure}

The formation of SMBHBs is more likely in galaxy clusters. The nearest galaxy cluster to Earth is the Virgo cluster. In Figure \ref{fig:VIRGO} we examine the possibilities for pulsar timing to detect GWs generated by SMBHBs in the Virgo cluster. The mean sky-position of this cluster is at a right ascension of 12$^\rmn{h}$30$^\rmn{m}$ and a declination of +12$^\circ$ \citep{2007ApJ...655..144M}; to produce this sensitivity curve all simulated GW signals come from this direction. The plotted sensitivity curve indicates that, with a false alarm probability of 1\%,  we have a better than 95\% probability of detecting sinusoidal signals in our timing residuals caused by $10^{10}\,\rmn{M}_\odot - 10^{10}\,\rmn{M}_\odot$ SMBHBs in the Virgo cluster with any polarisation at a range of frequencies and marginally also some $10^9\,\rmn{M}_\odot - 10^9\,\rmn{M}_\odot$ SMBHBs.

\begin{figure*}
\centering
\epsfig{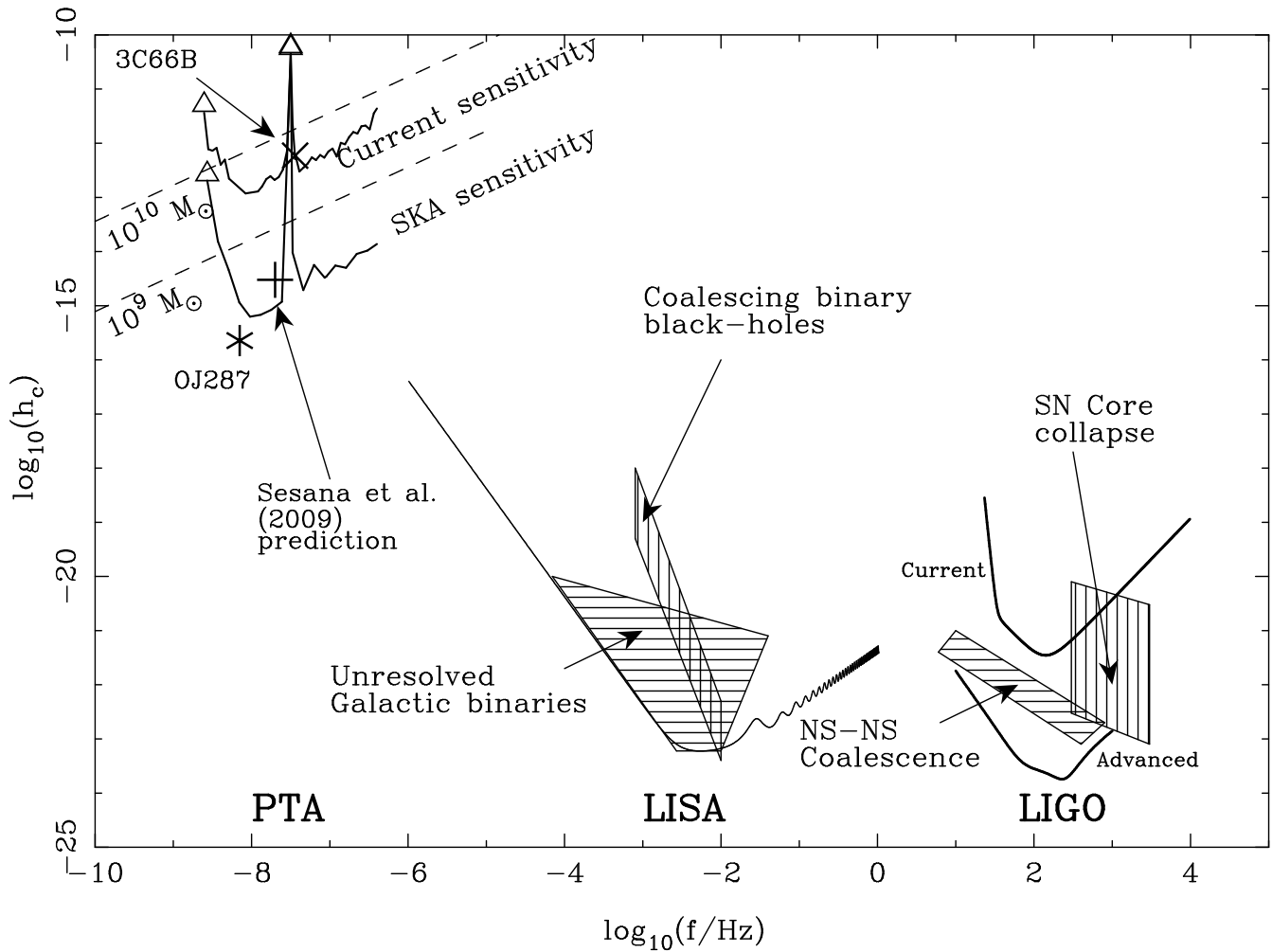}
\caption{
Sensitivity of some current and future GW observatories to individual GW sources as a function of frequency. The abscissa gives the GW frequency, the ordinate gives the minimum detectable strain amplitude of a sinusoidal GW point source with a random polarisation, phase and sky-position. For the pulsar timing array sensitivity we assume an incoherent detection scheme is used and the GW source position is unknown. The open triangles indicate that the plotted sensitivity at that frequency is a lower bound. The plot also shows some potentially detectable sources in the three frequency bands. The straight lines indicate the expected signals from two different types of SMBHB if they were located in the Virgo cluster, with equal member masses $10^{9}$M$_\odot$ and $10^{10}$M$_\odot$ as labelled. The `$\times$' symbol is the expected signal at the Earth caused by the proposed SMBHB at the core of the radio galaxy 3C66B. The `$*$' symbol is the expected signal caused by the candidate SMBHB at the core of OJ287. The `+' symbol is the GW strain and frequency emitted by a typical resolvable SMBHB as plotted in Figure 2 of \citet{2009MNRAS.394.2255S}. ``Unresolved galactic binaries'' include white-dwarf and neutron-star binaries. ``Coalescing binary black holes'' show the expected range of signals from the final inspiral of black-hole binary systems. The ``Current'' LIGO sensitivity shows the capabilities of existing datasets, while ``Advanced'' LIGO expects to improve GW sensitivity by two orders of magnitude. ``SN [supernova] core collapse'' and ``NS-NS [neutron star] coalescence'' are typical signals that LIGO expects to detect.
}
\label{fig:PSR_LISA_LIGO}
\end{figure*}

The PPTA sensitivity is complementary in GW frequency to the LIGO, VIRGO and LISA sensitivities. In Figure \ref{fig:PSR_LISA_LIGO} we give the detection sensitivity of some current and future GW detection experiments. Also shown on the plot are some likely sources in each of the detectable bands. This sensitivity curve now almost covers the full GW frequency range from $\sim$nHz through to $\sim$mHz; this frequency coverage will enable the study of the evolution of GW-emitting systems.

To obtain the LISA sensitivity curve, we have assumed the standard parameters for the LISA design and that it aims to detect sources at a signal-to-noise ratio of three. The LIGO sensitivity curves are obtained from the stated design goals of the project.

\subsection{The implied constraint on the merger rate of SMBHBs}

Non-detection of single-source GWs in the Verbiest et al. data enables an upper limit to be placed on the rate of super-massive black hole mergers \citep{Wen09}. To use the techniques presented in \cite{Wen09}, it is necessary to calculate the limiting sensitivity of our array at a matrix of GW frequency and strain values. The frequency values chosen were 50 logarithmically-spaced frequencies between $\left(30\,\rmn{years} \right) ^{-1}$ and $\left( 4\,\rmn{weeks}\right)^{-1}$, while the strain values were 50 logarithmically-spaced amplitudes between $10^{-16}$ and $10^{-10}$. The sensitivity at a frequency of (1\,yr)$^{-1}$ was also calculated, resulting in 51 frequency values overall. 1000 Monte Carlo iterations were used at each value of GW frequency and strain. The sensitivity matrix obtained gives a 95\% confidence contour which is consistent with the 95\% confidence upper bound obtained earlier (the thick dotted line in Figure \ref{fig:SensCurve}).

\begin{figure*}
\centering
\epsfig{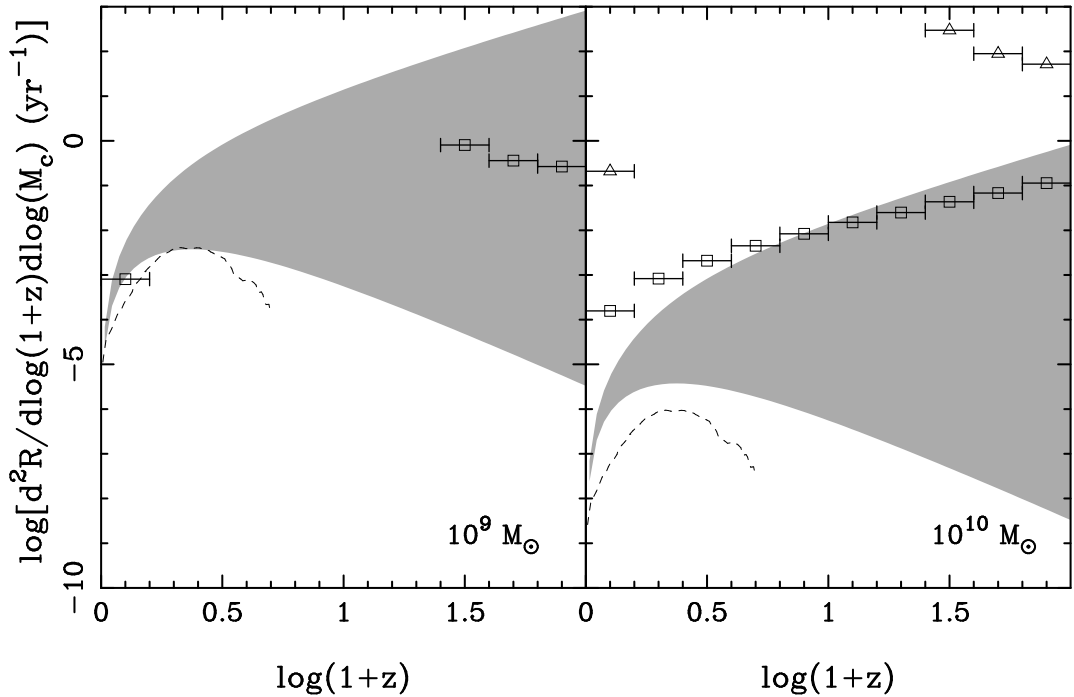}
\caption{ 
Upper limit on the rate of SMBHB mergers as a function of redshift (with a range of chirp masses) given the non-detection of any GW sources in the Verbiest et al. datasets. The open triangles give the upper limit on the SMBHB merger rate for the Verbiest et al. dataset and the open squares give the limit for the simulated SKA datasets. The shaded region indicates the expected coalescence rate obtained from \citet{2003ApJ...583..616J} as well as data from the Sloan Digital Sky Survey \citep{2009ApJ...692..511W} for SMBHB systems of chirp mass as labelled in each panel. The dashed line indicates the merger rate based on \citet{2008MNRAS.390..192S}.
}
\label{fig:rate}
\end{figure*}

This sensitivity matrix is used to provide an upper limit on the differential rate of SMBHB coalescence per logarithmic redshift and chirp mass. The results of this analysis are shown in Figure \ref{fig:rate}. Our data do not yet constrain the merging frameworks discussed by \cite{2003ApJ...583..616J} or \citep{2008MNRAS.390..192S} at the range of chirp masses we have considered. However, in coming years some of the high-mass and high-redshift predictions may be ruled out or confirmed using pulsar timing.

\subsection{A predicted SKA sensitivity curve}

Figure \ref{fig:PSR_LISA_LIGO} also gives a predicted sensitivity curve for the Square Kilometre Array (SKA)\footnote{See http://www.skatelescope.org/}. To produce this figure we chose 100 pulsars from the Australia Telescope National Facility pulsar catalogue \citep{mhth05}. We have assumed we can time each pulsar with an accuracy of 20\,ns over five years, obtaining one timing point per pulsar every two weeks. We have also assumed that their power spectra will be statistically white. This means the plotted sensitivity is a lower bound on what is achievable with the SKA for the assumed parameters, especially at low frequencies where we expect higher noise levels caused by the stochastic background of GWs and intrinsic pulsar timing noise.

The simulated SKA data are regularly sampled with equal error bars, which means that the level of spectral leakage will be much lower than that observed in irregularly sampled datasets with highly variable error bars. This means the confusion between red noise and low-frequency signal is no longer an issue in these simulations because a sinusoidal GW signal will induce a very narrow peak in each pulsar's power spectrum, even at low frequencies. We have therefore modelled each pulsar power spectrum with a constant.

There are three prominent losses in sensitivity - at frequencies smaller than $\left(T_\rmn{obs}\right)^{-1}$ and at periods of one year and six months. The partial loss in sensitivity at a period of six months ($\sim 6\times10^{-8}$\,Hz) is caused by fitting for the pulsar parallax. The total loss in sensitivity at GW periods of one year could be mitigated using independent measurements of the position of the pulsar, for example using very-high-precision interferometry; such precision may be available in the SKA era.

The SKA sensitivity curve shown in Figure \ref{fig:PSR_LISA_LIGO} is calculated assuming we do not know the location or frequency of a potential GW source; using these two additional pieces of information it will be possible to confirm or deny the binarity of the massive dark object at the core of OJ287, as well as resolve many of the SMBHBs predicted by \citet{2009MNRAS.394.2255S}. Using the SKA and LISA, it will also be possible to observe the full evolution of some \mbox{SMBHBs} from emitting GWs in the pulsar timing band (during the early phases of coalescence) to emitting GWs in the LISA band (during coalescence) \citep{2008JPhCS.122a2004P}.

\section{Conclusion} \label{sec:conc}

We have presented the strain sensitivity of the Parkes Pulsar Timing Array to sinusoidal point sources of GWs as a function of frequency. The sources most likely to produce a detectable sinusoid in the pulsar timing frequency range are super-massive black-hole binary systems in the early phases of coalescence at the cores of merged galaxies. The sensitivity curve is analogous to the LIGO, VIRGO and LISA sensitivity curves and indicates the unique GW frequency range accessible with pulsar timing. These results can be used to place an upper bound on the number of coalescing binary systems of a given chirp mass as a function of redshift. Current observations do not yet rule out any likely GW sources.

\section*{Acknowledgments}

This work is undertaken as part of the Parkes Pulsar Timing Array project. The Parkes radio telescope is part of the Australia Telescope, which is funded by the Commonwealth of Australia for operation as a National Facility managed by the Commonwealth Scientific and Industrial Research Organisation (CSIRO). This research was funded in part by the National Science Foundation (grant \#0545837) and RNMÕs Australian Research Council Federation Fellowship (project \#FF0348478). DRBY is funded by an APA and the CSIRO OCE PhD scholarship program. GH is the recipient of an Australian Research Council QEII Fellowship (\#DP0878388). JPWV is supported by a WVEPSCoR research challenge grant held by the WVU Center for Astrophysics.

\bibliographystyle{mn2e}
\bibliography{/Users/yar016/Desktop/bib_files/psrrefs,/Users/yar016/Desktop/bib_files/mybibliography,/Users/yar016/Desktop/bib_files/modrefs,/Users/yar016/Desktop/bib_files/crossrefs}

\section*{Appendix: Details of the Detection Technique}

In this section we give a detailed description of our detection technique, in particular describing some of the problems that arose during this treatment.

\subsection*{Our Technique for Producing a Sensitivity Curve}

Our method for creating curves showing the sensitivity of our timing residuals to GW-induced sinusoidal signals from individual SMBHBs takes into account the non-Gaussian noise which is a feature of many timing residual datasets. To produce a sensitivity curve for a given set of pulsars and their timing residuals, we use a 3-step process as follows:

\begin{enumerate}

\item \label{item:0} We choose logarithmically spaced GW frequencies between $\frac{1}{T_\rmn{obs} }$ and $\frac{N_\rmn{pts}}{2T_\rmn{obs}}$ (single pulsar) or between $(30\,\rmn{years})^{-1}$ and $(4\,\rmn{weeks})^{-1}$ (multiple pulsars). The frequency sampling we used for multiple pulsars requires over-sampling each power spectrum by a factor $30\,\rmn{yr} / T_\rmn{obs}$ for that pulsar.

\item At each frequency, we:   

\begin{enumerate}
  
\item \label{item:1} add the effect of a sinusoidal GW point source with angular frequency $2\pi f_i$, amplitude $h_{s}$ and random sky-position and polarisation to the TOAs, as described in equation (\ref{eq:ampres}).

\item\label{item:fit} process the data using the \textsc{Tempo2} pulsar timing software to obtain post-fit timing residuals.

\item \label{item:2} run a detection algorithm (described below) on the post-fit residuals which reports either a detection or a non-detection.

\item repeat steps \ref{item:1} -- \ref{item:2} a large number of times (we use 1 000 iterations) and record the detection percentage. 

\item If we have detected $\left(95\pm1\right)$ \% of the signals then we have satisfied our detection criterion and we record $f_i$ and $h_{s}$, which places a point on the pulsar timing sensitivity curve. If this criterion is not satisfied, adjust $h_{s}$ higher if too few detections have been made and lower if too many, then return to step \ref{item:1}.
\end{enumerate}

\item Select the next frequency in the grid and repeat.

\end{enumerate}

Our detector functions as follows:
\begin{enumerate}

\item For each pulsar in the input data, we calculate a power spectrum of the residuals using a Lomb-Scargle periodogram, with the frequency range described above.

\item We smooth the power spectrum by taking the logarithm of the power values and using a boxcar median filter (by default the number of points in the filter is 11 times the oversampling factor for that pulsar, in order to account for the correlated spectral estimates induced both by oversampling and by the irregular time sampling of the timing residuals).

\item \label{item:lsq} We use a least-squares fit to obtain a low-order polynomial (i.e. of order less than six) that provides a simple model of the median-smoothed log-spectrum. The median-smoothing and model-fitting are performed only on those points for which the frequency is $ \geq \left(T_\rmn{obs}\right)^{-1}$. This three-step spectral modelling process ensures that the simulated GW source is not included in the model as part of the noise in the spectrum; this is particularly important at the low- and high-frequency edges of the power spectrum. When analysing the data collected from multiple pulsars we combine their power spectra using a weighted sum. The weight used for each pulsar is the inverse of the simple frequency-dependent model of the power spectrum for that pulsar.

\item We multiply the noise model obtained above by a factor of $\sim$2-3 determined from simulation (see below) to define a set of detection thresholds for any given false alarm probability (we use $P_f = 1\%$). These detection thresholds are set such that the probability of any observed power across the whole spectrum being greater than the threshold when there is no signal present is 1\%.

\item If the measured power in the channel containing the input GW frequency is greater than the detection threshold in that channel, then we have made a detection of a significant sinusoid. 

\end{enumerate}

Some of the simulated sinusoidal GW point sources produce large signals in the timing residuals, depending on their polarisation and location on the sky. If a set of timing residuals showed evidence of a strong signal, a typical analysis would use a model of the pulsar with the fewest possible parameters (i.e. a period, period-derivative and any arbitrary phase offsets) to obtain residuals and then examine the dataset more closely. To simulate this process, in step \ref{item:fit} above, we calculate the full parameter fit as normal, but if the reduced-$\chi^2$ is larger than 20, then we instead only fit for the pulsar period, spin-down and jumps between datasets.

\subsection*{Our Technique for Producing an Upper Limit}

As described in Section \ref{sec:tech}, our technique for ruling out GWs with a particular strain amplitude as a function of frequency is much more straightforward than attempting to make a detection of the sinusoid induced by GWs emanating from SMBHBs. The important assumption in producing a correct upper limit without assuming anything about the statistics of the data in question is that, at any frequency in our power spectrum, the power caused by GWs cannot be {\it more} than the observed power; otherwise, we would have observed a higher power level at that frequency. That is, we assume that all the power at a given frequency is caused by GWs and then calculate the GW strain which gives a power greater than this level 95\% of the time.

To produce this limit, we first calculate the power spectrum of the observed timing residuals of each pulsar using the Lomb-Scargle periodogram. We then make a simple polynomial model of the noise in this spectrum and use the inverse of this noise model as the weight in calculating a weighted sum of the power spectra. This weighted and summed spectrum is the detection threshold. We then simulate noiseless GW signals (which are manifested as a pure sinusoid in each dataset), fit out as many of the pulsar parameters as possible from each sinusoid and calculate the same weighted sum described above (that is, using the noise model calculated for the observed timing residuals). Comparing this weighted sum of sinusoids to the detection threshold, we can scale the strain amplitude until we can detect the signal in 95\% of detection attempts. We can then rule out the existence of any stronger GW sources at this frequency (with random sky position and polarisation) with 95\% confidence.

\subsection*{The False Alarm Probability} \label{sec:fap}

We used simulation to calculate the correct detection threshold for a given dataset and a false alarm probability of 1\% across the whole spectrum. Any detection made will thus be a 3-$\sigma$ detection. The statistics of each channel in the power spectrum approximately follow a $\chi^2$-distribution, but many other issues change the statistics of each channel, as described below.

As soon as we add a large GW signal to our data in channel $i$, the statistics of channel $i$ follow a non-central $\chi^2$-distribution or a Ricean distribution. This does not affect the false-alarm probablility determination but would affect analytical determinations of pulsar timing sensitivity.

Other effects that change the statistics of each spectral channel include the irregular sampling of the time series (which can cause correlated estimates of the power at some frequencies), the oversampling of the power spectrum when analysing multiple pulsars (which means that the peaks in the power spectrum will be more fully resolved and thus the peak value is higher) and the median filtering (which lowers the height of each peak in the spectrum as well as raising the troughs).

Our method for calculating the height of the 3-$\sigma$ detection threshold was to simulate many realisations of white noise with an rms of 100\,ns and the same sampling as the original time series. Then, without performing any of the pulsar parameter fits or adding the effect of a SMBHB, calculate the average detection rate for {\it any} peak in the power spectrum to be greater than some estimated detection threshold. To perform the ``detection'' in this case we simply find the mean of the power spectrum (since the timing residuals are consistent with white noise) and then make the estimated threshold a factor of $\sim2$ higher than this mean. This factor is adjusted until the average detection rate for detections being made anywhere in the observed power spectrum equals the false alarm probability. In general the detection threshold had to be set at a factor of 1.3$-$2.5 higher than the threshold implied by assuming that each spectral channel follows a $\chi^2$-distribution.

\subsection*{Modelling the Power Spectrum} \label{sec:mods}

\begin{figure*}
\centering
\epsfig{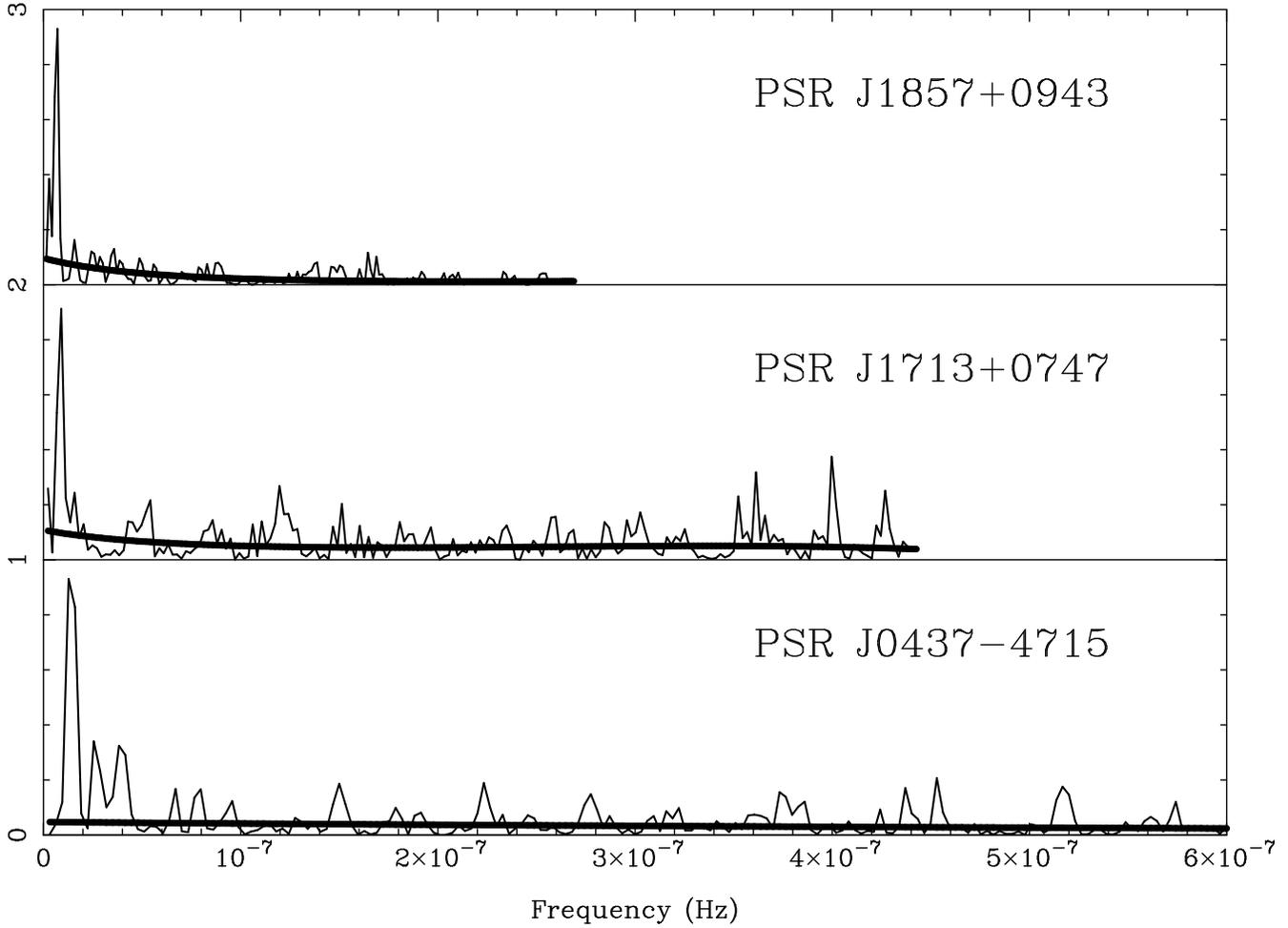}
\caption{
A typical power spectrum of each of three timing residual datasets, where we have added a large amplitude low-frequency source to each set of timing residuals. The abscissa gives the frequency examined, the ordinate gives the power in arbitrary units including constant offsets applied to the spectra of PSRs J1857+0943 and J1713+0747 to separate the spectra in making this plot. The thin trace is the power spectrum, the thick dark line is the adopted model for this spectrum. The low-order polynomial modelling accounts for the confusion between red noise in the timing residuals and signal leakage caused by irregular sampling. The power spectra of PSRs J0437$-$4715 and J1713+0747 have been modelled with quartics, while the spectrum of PSR J1857+0943 has been modelled with a cubic. The frequency coverage of PSR J0437$-$4715 extends to much higher frequencies than those shown because of the very large number of timing residuals for this pulsar.
}
\label{fig:models}
\end{figure*}

In Figure \ref{fig:models} we show a sample dataset with a very low frequency GW source injected and the models used for the three individual pulsars whose sensitivity is displayed in Figure \ref{fig:SensCurve}. In general, the models chosen are conservative in the presence of red noise to minimise the number of spurious detections at low frequencies. These figures demonstrate some general features of the power spectral models used. In particular, the models account for the varying levels of red noise and the possibility of signal leakage. When limiting the amplitude of the single sources that could be present in our data, we do not add sinusoids to the measured timing residuals and so a different model for the power spectrum may be used because the spectral features are different.

\label{lastpage}

\end{document}